\documentclass[conference]{IEEEtran}
\IEEEoverridecommandlockouts
 
\usepackage{cite}
\usepackage{amsmath,amssymb,amsfonts}
\usepackage{algorithmic}
\usepackage{graphicx}
\usepackage{textcomp}
\usepackage{xcolor} 
\usepackage{epsfig}
\usepackage[font=small,skip=0pt]{caption}
\usepackage{amsmath,amssymb,amsthm}
\usepackage{comment} 
\usepackage{tabularx}
\usepackage{algorithmic}
\usepackage{graphicx}
\usepackage{algorithm}
\usepackage{enumitem}
\usepackage{eufrak}
\usepackage{lipsum}
\usepackage{stfloats}
\usepackage{bbm}
\usepackage{relsize}
\usepackage{multicol}

\newtheorem{thm}{Theorem}
\newtheorem{lem}{Lemma}

\def\bsum{\mathlarger{\sum}}

\def\BibTeX{{\rm B\kern-.05em{\sc i\kern-.025em b}\kern-.08em
    T\kern-.1667em\lower.7ex\hbox{E}\kern-.125emX}}
\begin{document}

\title{  Optimal Bandwidth Allocation for Multicast-Cache-Aided on-Demand Streaming in Wireless Networks }

\author{
  \IEEEauthorblockN{Mohsen Amidzadeh$^*$, Olav
          Tirkkonen$^*$ and Giuseppe Caire$^\dag$}
	\IEEEauthorblockA{$^*$Department of Communications and
          Networking, Aalto University, Espoo, Finland\\
          $^\dag$Communications and Information Theory Chair, TU Berlin, Germany\\
	\{mohsen.amidzade, olav.tirkkonen\}@aalto.fi, giuseppe.caire@gmail.com}	
}
\maketitle

%----------------------Abstract------------------
\begin{abstract}
We consider a hybrid delivery scheme for streaming content, combining
cache-enabled Orthogonal Multipoint Multicast (OMPMC) and on-demand
Single-Point Unicast (SPUC) transmissions for heterogeneous networks.
The OMPMC service transmits cached files through the whole network to
interested users, and users not being satisfied by this service are
assigned to the SPUC service to be individually served. The SPUC
fetches the requested files from the core network and unicasts them to
UEs using cellular beamforming transmissions. We optimize the delivery
scheme to minimize the average resource consumption in the network.
We formulate a constrained optimization problem over the cache
placement and resource allocation of the OMPMC component, as well as the
multi-user beamforming scheme of the SPUC component. We apply a
path-following method to find the optimal traffic offloading solution.
The solutions portray a contrast between the total amount of consumed resources
and service outage probability.
Simulation results show that the hybrid scheme provides
a better tradeoff between the amount of network-wide consumed
resources and the service outage probability, as compared to schemes
from the literature.

\end{abstract}

\begin{IEEEkeywords}
Hybrid content delivery, wireless caching, multipoint multicasting, single-point unicasting, 
resource consumption, zero-forcing beamforming, parametric optimization.
\end{IEEEkeywords}

%/\/\/\/\/\/\/\/\/\/\/\/\/\/\/\/\/\/\/\/\/\/\/\/\/\/\/\/\/\/\/\/\/\/\/\/\/\/\/\/\/\/\/\/\/\/\/\/\/\/\/\/\/\/\/\/%
%  													   	 Introductoin    										%
%/\/\/\/\/\/\/\/\/\/\/\/\/\/\/\/\/\/\/\/\/\/\/\/\/\/\/\/\/\/\/\/\/\/\/\/\/\/\/\/\/\/\/\/\/\/\/\/\/\/\/\/\/\/\/\/%
\section{Introduction}\label{Sec_Intro}

Wireless caching~\cite{Bastug2014} is a candidate method to alleviate
the unprecedented data congestion and traffic escalation issues in
cellular networks. To determine a cache policy, the two phases of cache
placement and cache delivery have to be defined~\cite{Liu2016}, and
both should be optimized to achieve a viable cache strategy.

Content placement can be performed using a probabilistic
\cite{Blaszczyszyn2015,Wu2017,Cheng2019}, or a deterministic approach
\cite{Liu2016,Zhao2018,Zhou_2018}, where the probabilistic one can be
applied to large networks in a scalable manner. In
\cite{Blaszczyszyn2015}, probabilistic content placement was
proposed by which the cache-equipped nodes independently and randomly
store files based on a probability distribution.
%In \cite{Wu2017,Chen2017}, a tier-specific probabilistic caching was exploited for the Heterogeneous Networks (HetNets), 
%in which tier $n$ cache file $m$ with a specific probability.
In \cite{Cheng2019}, a two-tier Heterogeneous Network (HetNet) was
considered with a hybrid cache placement strategy, combining
deterministic caching in one tier
%$1$-tier,
and probabilistic caching in the other.
%$2$-tier, was devised.

For the
%In the context of
content delivery,
%policies,
multipoint multicast
(MPMC), single-point multicast (SPMC), and single-point unicast (SPUC)
transmission schemes should be distinguished. MPMC utilizes
multiple serving nodes to cooperatively broadcast files through the
whole network, whereas the SPUC exploits on-demand transmissions to
individually satisfy requesting User Equipments (UEs). For the SPMC
scheme, each caching Base Station (BS) 
%is employed by the network to
multicasts a file to several requesting UEs.

SPUC has been considered in \cite{Wu2017,Zhao2018,Wu2020} as an
on-demand cache delivery scheme for HetNets. In \cite{Wu2017}, 
%HetNet with
Zero Forcing Beamforming (ZFB) BSs and cache-equipped
helper nodes were considered where the requesting UE is associated with
the helper node with maximum received power.
%In \cite{Li2018}, SPUC was considered for a HetNet where each UE is associated 
%with the tier whose nearest BS has maximum received power.
In \cite{Wu2020},
%a tier-level
random resource allocation was utilized with SPUC.
%to optimize a cache policy from the
The fraction of bandwidth and the probabilistic cache placement were
optimized to maximize offloading.

SPMC has been used in \cite{Cui2017, Zhong2018, Ye2019} as a content
delivery method. In \cite{Cui2017, Ye2019}, probabilistic caching is
exploited for a two-tier HetNet where each BS multicasts $k$ files in
disjoint resources.
% each spanning $1/k$ of total bandwidth.
In \cite{Zhong2018}, coded caching is applied for a cellular network
with multi-antenna BSs. The authors exploit joint SPUC and SPMC
beamforming for content delivery, and the beamforming vectors are
optimized to maximize the minimum rate of the UEs. 

Digital Terrestrial TV Broadcasting systems~\cite{Sari1995} 
deliver content based on multipoint broadcast transmissions, while
multipoint multicast (MPMC) delivery is used in the Long Term
Evolution (LTE) system in the context of Multimedia Broadcast Multicast Services
%MBMS
\cite{Wibowo2011}. In a multi-cell transmission mode, all serving BSs
stream the same file through the whole network in a
Single-Frequency-Network (SFN) configuration.
%Therefore, each file is concurrently transmitted over the same frequency bandwidth.
A SFN-configured MPMC was utilized in \cite{Ours_2020} for edge
caching cellular networks. An orthogonal MPMC (OMPMC) was considered
where resources are network-wide orthogonalized among cached files.
The cache policy was then designed with respect to (w.r.t)
file-specific resource allocation and probabilistic cache placement.

In this paper, we devise a hybrid content delivery scheme for
cache-enabled multiantenna HetNets, combining OMPMC and SPUC. SPUC can serve
individual users, but 
%For networks
%with multiple UEs requesting files in the same frequency bandwidth,
%SPUC
suffers from Co-Channel Interference (CCI). In contrast, OMPMC can serve a
population of users interested in a given file with limited CCI~\cite{Ours_2021}.
%This issue can be coped
%with multicast transmission towards UEs that are receiving files in
%the same frequency bandwidth.
However, this is obtained at the cost of the increase in the outage
probability, as the multicast transmission can not individually
satisfy UEs according to the resources need for successful reception.
Also, if only few users are interested in a file, using OMPMC would
consume large amounts of network resources as compared to SPUC. This
portrays a trade-off between OMPMC and SPUC delivery. We exploit this
to develop a hybrid scheme based on OMPMC that streams the files using
a network-wide file-specific resource allocation, and on-demand SPUC
that unicasts the files using UE-specific resource-allocations. In
contrast to \cite{Zhong2018,Ours_2022}, we devise the hybrid scheme
such that all UEs being dissatisfied by the multicast component can be
properly served by the network.
%Further, in \cite{Ours_2022} the overall outage probability of the network was optimized 
%regardless of the resource consumed by the whole network.
We optimize the network w.r.t. total amount of consumed resources
subject to inevitable outages due to coping with the possibility of
too large bandwidth requests.
%To get an analytic handle on the problem, we model the large network using PPP for both UEs and BSs. 

More specifically,
%we present a hybrid delivery scheme combining SPUC
%and OMPMC transmissions for traffic offloading policy in edge caching
%heterogeneous networks. We
for the proposed hybrid SPUC/OMPMC delivery scheme,
we  find expressions for the outage
probability and  network resource consumption,
%of the whole network as the system performances,
using stochastic geometry.
%, and draw some intuitions based on these expressions.
We then formulate the optimal traffic offloading problem as a joint
optimization problem over cache placement, resource allocation and
multiuser beamforming. To solve this problem, we represent it as a
parametric optimization problem and leverage a path-following method
to find the global optimal policy from the perspective of total
resource consumption. We compare the proposed hybrid scheme to cache
delivery policies from the literature, in different settings.
	
The remainder of this paper is organized as follows. In Section
\ref{Sec_Model}, the system model is introduced. In
Section~\ref{Sec_TotalResource}, the total resource consumption of the
considered hybrid scheme is computed. The optimization problem is
formulated in Section \ref{Sec_Optimizatoin}, while simulation results
are presented and discussed in Section \ref{Sec_Result}. Finally,
Section \ref{Sec_Conclu} concludes the paper.

\textbf{Notations}: In this paper, we use lower-case $a$ for scalars, 
bold-face lower-case $\textbf{a}$ for vectors 
and bold-face uppercase $\textbf{A}$ for matrices. 
Further, $\{a_n\}_1^N$ collects the components of vector $\boldsymbol{a}$ from $n=1$ to $n=N$.
We use $\textbf{1}$ and $\textbf{0}$ to denote the vector with all elements equal to one 
and zero, respectively.
We use $\dot{\boldsymbol{a}}(\theta )$ to represent the derivative of $\boldsymbol{a}(\theta)$ 
with respect to $\theta$.

%/\/\/\/\/\/\/\/\/\/\/\/\/\/\/\/\/\/\/\/\/\/\/\/\/\/\/\/\/\/\/\/\/\/\/\/\/\/\/\/\/\/\/\/\/\/\/\/\/\/\/\/\/\/\/\/%
%  													   	 System Model  											%
%/\/\/\/\/\/\/\/\/\/\/\/\/\/\/\/\/\/\/\/\/\/\/\/\/\/\/\/\/\/\/\/\/\/\/\/\/\/\/\/\/\/\/\/\/\/\/\/\/\/\/\/\/\/\/\/%
\section{System Model} \label{Sec_Model}

We consider a content library containing $N$ different files. 
%There is a set of active UEs, and in a given time slot, each UE requests one
%of the files.
%popularity $\{f_n\}_1^N$, where
The fraction of active UEs requesting file $n$ is $f_n$. We assume
that file popularities $\{f_n\}_1^N$ are known to the network. We use
the Zipf distribution (see \cite{Breslau1999}) to model popularity;
$f_n = n^{-\tau}/\sum_{m=1}^N m^{-\tau}$, with $\tau$ being the
skewness of the Zipf distribution. We consider a streaming
service and for simplicity, we assume the same streaming rate
$R$ for all files which leads to the identical service quality. The
study of dynamic rate/quality is left for future work.
%files have the same size equal to one.

We consider a two-tier HetNet with BSs and Helper Nodes (HNs).
The network applies a hybrid delivery scheme, based on OMPMC and SPUC
components to serve UEs. The HNs constitute the multicast
component and are equipped with limited-capacity caches which
proactively cache files based on a probabilistic
approach~\cite{Serbetci2017}. The HNs apply OMPMC to broadcast the cached
files
%. More specifically, HNs cooperatively stream the cached contents
through the whole network in file-specific disjoint resources. In the
OMPMC layer, there may be co-channel interference arising from
far-away HNs. However, with conventional Orthogonal Frequency Division
Multiplexing (OFDM) parameterization, such co-channel interference
would be negligible~\cite{Ours_2021}. 
%Accordingly, there will not be any co-channel interference during
%multicast delivery service.
The BSs constitute the unicast component. They are connected to a
high-capacity backhaul link which can fetch files from the
core-network, and serve these to users using on-demand unicast. The
BSs are equipped with $L$ antennas and use multiuser ZFB 
to serve  $u$ users simultaneously in a frequency resource.

The network dedicates network-wide disjoint resources for the OMPMC
and SPUC delivery components, denoted by $W^{\rm MC}$ and $W^{\rm
  UC}$, respectively. Each BS uses the full bandwidth $W^{\rm
  UC}$, i.e., frequency reuse 1 is used in the unicast layer. 
%This is better in the content delivery subsec
%Any file request not being satisfied by the
%multicast component is transferred to the unicast component.
When served by the unicast layer, a UE is connected to the nearest BS.
To support the streaming rate $R$, 
the BS allocates a sufficient amount of resources to a UE, if it is
not greater than a service threshold.

We assume that the locations of UEs, BSs and HNs are
based on three independent Poisson Point Processes (PPPs), i.e., $\Phi_u$ and $\Phi_b$ and
$\Phi_h$, respectively,  with intensities 
$\lambda_u$, $\lambda_b$ and $\lambda_h$.

%/\/\/\/\/\/\/\/\/\/\/\/\/\/\/\/\/\/\/\/\/\/\/\/\/\/\/\/\/\/\/\/\/\/\/\/\/\/\/\/\/\/\/\/\/\/\/\/\/\/\/\/\/\/\/\/%
%  													   	 SubSec_Placement										%
%/\/\/\/\/\/\/\/\/\/\/\/\/\/\/\/\/\/\/\/\/\/\/\/\/\/\/\/\/\/\/\/\/\/\/\/\/\/\/\/\/\/\/\/\/\/\/\/\/\/\/\/\/\/\/\/%
\subsection{Cache Placement and Content Fetching}
\label{SubSec_Placement}

The HNs are equipped with caches of limited memory
capacity so that they can store at most $M$ files. They proactively 
and independently cache files based on a probabilistic
approach \cite{Blaszczyszyn2015} modeled by a common probability distribution.
As such, cache weights $\{p_n\}_1^N$ with $0 \leq p_n \leq 1$ and 
$\sum_{n=1}^N p_n  \leq M$ are introduced, and an approach is devised
such that each HN can store exactly $M$ files with file $n$ being 
cached at the HN with probability equal to $p_n$. 
Note that the files are cached at HNs based on another PPP with intensity
$\{ \lambda_u p_n\}_1^N$ , according to the thinning property of $\Phi_u$ \cite{BK1}.

The BSs providing on-demand unicast delivery reactively fetch the
requested files from the core network if they are requested by UEs.

%/\/\/\/\/\/\/\/\/\/\/\/\/\/\/\/\/\/\/\/\/\/\/\/\/\/\/\/\/\/\/\/\/\/\/\/\/\/\/\/\/\/\/\/\/\/\/\/\/\/\/\/\/\/\/\/%
%  													   	 SubSec_Delivery										%
%/\/\/\/\/\/\/\/\/\/\/\/\/\/\/\/\/\/\/\/\/\/\/\/\/\/\/\/\/\/\/\/\/\/\/\/\/\/\/\/\/\/\/\/\/\/\/\/\/\/\/\/\/\/\/\/%
\subsection{Content Delivery}\label{SubSec_Delivery}

The network operates in a time-slotted fashion. Within each time-slot,
requests arrive at arbitrary times from a spatial realization of UEs
based on $\Phi_u$. The network applies a hybrid delivery scheme based
on OMPMC and SPUC transmissions to satisfy these requests. Any UE not
being satisfied by the multicast component is assigned to the unicast
component to be properly served.
%Notice that the total resources of the unicast and multicast components are orthogonal to each other.

The multicast component streams cached files through whole network by
cooperation of all HNs. It uses file-specific disjoint resources for
broadcasting different files bandwidth $w_n^{\rm MC}$ allocated for
file $n$. To reduce the latency of streaming in the OMPMC mode, we
apply Harmonic Broadcasting (HB) characterized by harmonic number
$H_w$~\cite{Juhn1997}, though the analysis is not restricted to any
particular streaming scheme. When the consumed bandwidth is increased
by a factor of $H_w = \sum_{i=1}^D 1/i$, the average latency of a user
requesting the file within the time slot is reduced by a factor of
$1/D$.
%This number leads to a specific latency
%for UEs that should wait to receive their requested files from
%beginning.
We assume that the average transmission power of all HNs is
the same, denoted by $p_{\rm tx}$.
%Further, each BS allocates a fractional power $p_{\rm tx} {w_n^{\rm MC}}/{W^{\rm MC}}$ 
%to broadcast this file in resource $w_n^{\rm MC}$, where $W^{\rm MC}=\sum_{n=1}^N w_n^{\rm MC}$ 
%is the total transmission bandwidth for the multicast component.
With equal power allocation across frequency, the transmitting
Signal-to-Noise-Ratio (SNR) of all files in OMPMC are the same,
$  \gamma^{\rm MC}_{\rm tx} = \dfrac{p_{\rm tx}}{W^{\rm MC}N_0}
$,
where $N_0$ is the noise spectral density.
%Considering that it is independent of file index $n$, 
%hereafter we denote it by $\gamma_{\rm tx}^{\rm MC}$.

Any UE not being satisfied by the multicast component, requests the
file from the unicast component. As such, the BSs in the unicast layer,
constitute a Poisson-Voronoi tessellation with different cell sizes.
%It then assigns
The nearest BS to the UE
%transfers the file. The responsible BS
fetches the file from the core network and unicasts it 
towards the UE. If the Signal-to-Interference-plus-Noise Ratio (SINR)
of a user is above a threshold, the responsible BS responds to the UE by
allocating the resources it needs to successfully decode its files.
In each Voronoi cell, there may be multiple UEs
requesting files.

We assume that the average transmission power of all BSs is $p_{\rm tx}$, 
and they are equipped with $L$ antennas and
serve $u$ UEs at the same frequency bandwidth using a ZFB
transmission.

%/\/\/\/\/\/\/\/\/\/\/\/\/\/\/\/\/\/\/\/\/\/\/\/\/\/\/\/\/\/\/\/\/\/\/\/\/\/\/\/\/\/\/\/\/\/\/\/\/\/\/\/\/\/\/\/%
%  													   	 SubSec_ResourceMPMC									%
%/\/\/\/\/\/\/\/\/\/\/\/\/\/\/\/\/\/\/\/\/\/\/\/\/\/\/\/\/\/\/\/\/\/\/\/\/\/\/\/\/\/\/\/\/\/\/\/\/\/\/\/\/\/\/\/%
\subsection{Resource Consumption of Multicast Component} \label{SubSec_ResourceMPMC}
The multicast component applies network-wide resource allocation 
with OFDM-based transmission to broadcast the files through the network.
%there is not any co-channel interference  and
The Signal-to-Noise-Ratio (SNR) of UE $k$ requesting file $n$ is expressed as \cite{Ours_2020}:
\begin{equation} \label{EQ_SNR_MPMC}
	\gamma_{k,n}^{\rm MC}= \gamma_{\rm tx}^{\rm MC} \sum_{j\in\Phi_{p,n}}{ |h_{j,k}|^2 \|\boldsymbol{x}_k-\boldsymbol{r}_j \|^{-e}},
\end{equation}
where
%$\gamma_{\rm tx}^{\rm MC}=p_{\rm tx}/(W^{\rm MC}N_0)$,
$\Phi_{p,n}$ stands for the set of HNs caching file $n$, 
$h_{j,k}$ is the channel coefficient between HN $j$ and UE $k$, 
$\boldsymbol{x}_k$ and $\boldsymbol{r}_j$ are the locations of UE $k$ and HN $j$, respectively,
and $e$ is the path-loss exponent.
We use a standard distance-dependent to model the path-loss, 
and assume an Rayleigh distribution for the channel coefficient, i.e., $|h_{j,k}|^2\sim \exp(1)$.

The maximum achievable rate for a transmission is obtained
from
%based on the channel capacity
%at the presence of
the corresponding Additive-White-Gaussian-Noise channel capacity. If
the maximum rate experienced by a UE is less than the streaming rate
$R$, the UE is in outage. The outage probability
$\mathcal{O}_{n,k}^{\rm MC}$ for UE $k$ requesting file $n$ then is:
$$
\mathcal{O}_{n,k}^{\rm MC} = \mathbb{P}\{ w_n^{\rm MC} \log_2(1+\gamma_{k,n}^{\rm MC}) \leq R \}.
$$
Defining a spectral efficiency threshold $\alpha_n = R/w_n^{\rm MC}$,
the total resource usage of the multicast component is $W^{\rm
  MC}= \sum_{n=1}^N w_n^{\rm MC} = \sum_{n=1}^N \frac{R}{\alpha_n} $.
However, as we apply harmonic broadcast~\cite{Juhn1997}, the effective
resource usage will be multiplied by $H_w$, i.e.,
$$
W_{\rm eff}^{\rm MC}(\boldsymbol{\alpha}) = H_w \sum_{n=1}^N \frac{R}{\alpha_n}.
$$
Based on the Slivnyak-Mecke theorem \cite{BK1}, the performance can be computed 
for a typical UE located at the origin. 
Hence, we can set $\mathcal{O}_{n,0}^{\rm MC} = \mathcal{O}_{n}^{\rm MC}$.
Accordingly, the outage probability of file $n$, served by the multicast component, is \cite{Ours_2020}:
\begin{align}\label{EQ_out_MC}
\mathcal{O}^{\rm MC}_n\left(p_n,\{\alpha_l\}_l\right) &= \frac{2}{\pi} \int_0^\infty \Bigg\{ \frac{1}{w} 
\cos\Big(\frac{\pi^2\lambda_h p_n }{e \cos(\pi/e)} \Big(\frac{w}{g_{\boldsymbol{\alpha}}}\Big)^{2/e} \: \Big) \times \notag \\
& \exp\Big(\frac{-\pi^2\lambda_h p_n}{e \sin(\pi/e)} \Big(\frac{w}{g_{\boldsymbol{\alpha}}}\Big)^{2/e} \: \Big)  
\sin\bigg(\frac{w}{\gamma_R}  \bigg) \Bigg\}dw,
\end{align}
%and for $e=4$:
%$$
%\mathcal{O}^{\rm MC}_n\left(p_n,\{\alpha_l\}_l\right) = \erfc\Bigg( \dfrac{\pi^2}{4}\lambda_h p_n %\sqrt{\dfrac{\gamma_R}{g_{\boldsymbol{\alpha}}} }  \Bigg),
%$$
where $g_{\boldsymbol{\alpha}} = \big(2^{\alpha_n}-1\big)\sum_{l=1}^N \alpha_l^{-1}$ and $\gamma_R = \dfrac{p_{\rm tx}}{N_0}\dfrac{1}{R H_w}$.

%/\/\/\/\/\/\/\/\/\/\/\/\/\/\/\/\/\/\/\/\/\/\/\/\/\/\/\/\/\/\/\/\/\/\/\/\/\/\/\/\/\/\/\/\/\/\/\/\/\/\/\/\/\/\/\/%
%  													   	 SubSec_ResourceSPUC									%
%/\/\/\/\/\/\/\/\/\/\/\/\/\/\/\/\/\/\/\/\/\/\/\/\/\/\/\/\/\/\/\/\/\/\/\/\/\/\/\/\/\/\/\/\/\/\/\/\/\/\/\/\/\/\/\/%
\subsection{Resource Consumption of Unicast Component}
\label{SubSec_ResourceSPUC}

%To analyze the SINR of the unicast component, we consider that BSs
%constitute a Voronoi tessellation with different cells.
In each Voronoi cell of the unicast layer, zero-forcing
beamforming is utilized to serve $u$ UEs in the same bandwidth.
Consequently, for UE $k$ served by the unicast component, the SINR is:
\begin{align}\label{EQ_SINR}
\gamma_{k}^{\rm UC}=\frac{g_k \|\boldsymbol{x}_k - \boldsymbol{r}_0\|^{-e}}{ {1}/{\gamma_{\rm tx}^{\rm UC}} + \underbrace{\sum_{j\in\Phi_b \backslash \{0\}} g_j^k \|\boldsymbol{x}_k - \boldsymbol{r}_j\|^{-e}}_{I_k}}
\end{align}
where $\gamma_{\rm tx}^{\rm UC}=p_{\rm tx}/(W^{\rm UC}N_0)$,
$g_k$ is the effective channel gain between the nearest BS and UE $k$, 
constructed from the channel vector and the beamforming vector,
and $g_j^k$ is the effective channel gain from BS $j$ to UE $k$.
Further, $\boldsymbol{r}_0$ and $\boldsymbol{r}_j$ are the locations
of the nearest BS and BS $j$.
Consequently, we have $g_k \sim \Gamma(L-u+1,1)$ and $g_j^k \sim \Gamma(u,1)$ \cite{Chen2016},
with $\Gamma(a, b)$ being the gamma distribution with shape $a$ and scale $b$.

In each cell, the BS allocates a sufficient amount of resources to the
served UEs. However, if the SINR of a user is very small, near
infinite bandwidth would be needed to serve the user. To cope with this,
we apply a thresholding policy, the bandwidth allocated to serve  
UE $k$ is:
\begin{align}\label{EQ_serviceTh}
	w_k^{\rm UC} = \begin{cases}
	{R}/{\log_2(1+\gamma_{k}^{\rm UC})}, & \gamma_{k}^{\rm UC} \geq \gamma_{\rm th}\\
	0,  & {\rm otherwise}
\end{cases},
\end{align}
where $\gamma_{\rm th}$ is the SINR threshold. Hence, no resource is
allocated if the UE is in a poor condition. Considering that each BS 
applies multiuser ZFB towards $u$ users, the
resources consumed in each cell is approximately given by the total
amount of resources allocated to UEs in that cell divided by $u$. This
becomes precise when $\lambda_u/\lambda_b\to\infty$, and is a
sufficient approximation when $\lambda_u \gg \lambda_b$. We assume
that all BSs are active during unicast service, so the average
resource consumption by this component is determined by the average
consumption of a typical Voronoi cell $\mathcal{V}_0$:
\begin{align}\label{EQ_WUC0}
	W^{\rm UC}(\cdot) = \mathbb{E} \left\{ \sum_{k\in \mathcal{V}_0} \frac{w_k^{\rm UC}}{u} \right\}
\end{align}	
The expectation is w.r.t. Poisson processes of UEs ($\Phi_u$) and BSs
($\Phi_b$), as well as effective channel gains based on
\eqref{EQ_SINR}.

%/\/\/\/\/\/\/\/\/\/\/\/\/\/\/\/\/\/\/\/\/\/\/\/\/\/\/\/\/\/\/\/\/\/\/\/\/\/\/\/\/\/\/\/\/\/\/\/\/\/\/\/\/\/\/\/%
%  													   	 Sec_TotalResource										%
%/\/\/\/\/\/\/\/\/\/\/\/\/\/\/\/\/\/\/\/\/\/\/\/\/\/\/\/\/\/\/\/\/\/\/\/\/\/\/\/\/\/\/\/\/\/\/\/\/\/\/\/\/\/\/\/%
\section{Resource Consumption of Hybrid Scheme} \label{Sec_TotalResource}

For the resource consumption in the unicast layer we have 
\begin{thm}\label{Theorem_Wt}
%Suppose a Single-Point Unicast content delivery scheme
Consider an interference-limited frequency reuse 1 cellular network
with BSs and UEs coming from PPPs with intensities $\lambda_b$ and
$\lambda_u^{\rm eff}$, respectively, where UEs are served by their
nearest BS. BSs use ZFB with $L$ antennas towards $u$ simultaneous
users, and allocate bandwidth to users using service threshold
$\gamma_{\rm th}$ as in (\ref{EQ_serviceTh}).
%and based on their required
%bandwidth,
%and assuming all BSs are active during delivery service, 
The average amount of  resources needed is then:
\begin{align}\label{EQ_WUC}
\!\!\!\!\! W^{\rm UC}(u,\lambda_u^{\rm eff}) = \frac{\lambda_u^{\rm eff}}{2u\lambda_b } \sum_{l=1}^{L-u+1}f_{l,u} \int_0^{w_{\rm th}}
w \frac{d}{dw} C_{l,u}(w) dw,
\end{align}
where
\begin{align*}
	C_{l,u}(w) = \frac{1}{_2F_1\big(-\frac{2}{e},u,1-\frac{2}{e},-\xi\:l\:\eta(w)\big)},
\end{align*}
$w_{\rm th}={R}/{\log_2(1+\gamma_{\rm th})}$, 
$\eta(w) = 2^{R/w}-1$, 
$\xi= (L-u+1)!^{\frac{-1}{L-u+1}}$,
and $_2F_1(\cdot)$ is the hypergeometric function.
The outage probability of a UE
%being served with the SINR threshold $\gamma_{\rm th}$
is:
\begin{align*}
	\mathcal{O}^{\rm UC}(u,\gamma_{\rm th}) = 1-\sum_{l=1}^{L-u+1} \frac{f_{l,u}}{_2F_1(-\frac{2}{e},u,1-\frac{2}{e},-\xi\, l\, \gamma_{\rm th})},
\end{align*}
where $f_{l,u} = (-1)^{l+1}{L-u+1 \choose l}$. 
\end{thm}
\begin{proof}
	See appendix \ref{App1}.
\end{proof}
Notice that $W^{\rm UC}(\cdot)$ depends on ${\lambda_u^{\rm eff}}/{\lambda_b}$,
not on both $\lambda_b$ and $\lambda_u^{\rm eff}$ separately.
%Further, it is apparently seen that the dependency of $W^{\rm UC}(\cdot)$ on the reliable rate $R$ 
%is not linearly scalable, 
%nonetheless numerical evaluations show that it approximately behaves as a linear function w.r.t. $R$, 
%for different values of beamforming parameter $u$, path-loss exponent $e$ and number of antennas $L$.
%We also pay attention to the behavior of $W^{\rm UC}(\cdot)$ w.r.t $u \in \{1,\ldots,L\}$; 
%it decreases until reaching a minimum value and then increases, for different values of $e$ and $L$.

By comparing the results of Theorem \ref{Theorem_Wt}, 
we can relate the total resource $W^{\rm UC}$
to the outage probability $\mathcal{O}^{\rm UC}$:
\begin{align}	\label{EQ_W_Out}
	\frac{dW^{\rm UC}(u,\lambda_u^{\rm eff})}{d\gamma_{\rm th}}  = -\frac{w_{\rm th} \: \lambda_u^{\rm eff}}{2u \: \lambda_b}  \frac{d\mathcal{O}^{\rm UC}(u, \gamma_{\rm th})}{d\gamma_{\rm th}} 
\end{align}
\begin{comment}
\begin{align*}
& W^{\rm UC}(u,\lambda_u^{\rm eff}) = -\frac{\lambda_u}{2u \lambda_b} \int_0^{w_{\rm th}} w \frac{d}{dw} \mathcal{O}^{\rm UC}(u, \eta(w)) dw   \\
&\quad= \frac{\lambda_u}{2u \lambda_b} \Big( -w_{\rm th} \mathcal{O}^{\rm UC}(u,\gamma_{\rm th}) + \int_0^{w_{\rm th}} \mathcal{O}^{\rm UC}(u, \eta(w)) dw  \Big)
\end{align*}
For $\gamma_{\rm th}<1$, 
we have $  \int_0^{w_{\rm th}} \mathcal{O}^{\rm UC}(u, \eta(w)) dw\gg w_{\rm th} \mathcal{O}^{\rm UC}(u,\gamma_{\rm th})$,
which enables us to get the following approximation:
\begin{align}
\label{EQ_W_Out}
W^{\rm UC}(u,\lambda_u^{\rm eff}) \approx \frac{\lambda_u}{2u \lambda_b}  \int_0^{w_{\rm th}} \mathcal{O}^{\rm UC}(u, \eta(w)) dw 
\end{align}
\end{comment}
Equation \eqref{EQ_W_Out} directly provides a trade-off between the outage probability, 
and resource consumption of the unicast scheme.
Considering that $ \dfrac{d\mathcal{O}^{\rm UC}(u, \gamma_{\rm th})}{d\gamma_{\rm th}} > 0$, increase in $w_{\rm th}$ 
makes $W^{\rm UC}$ grows monotonically. 
%The less the service threshold is, the less the service outage is but the more resources are consumed.

Not all UEs request from the unicast component; only UEs being
dissatisfied by the multicast component do. Based on the thinning
property of PPP, $\lambda_{u}^{\rm eff}$ in \eqref{EQ_WUC} depends on
the overall outage probability of the multicast component: $
\mathcal{O}^{\rm MC}\left( \boldsymbol{p},\boldsymbol{\alpha}
\right)=\sum_{n=1}^N f_n \mathcal{O}_n^{\rm MC}\left(
p_n,\{\alpha_l\}_l \right), $ considering that $f_n$ is the popularity
of file $n$. Therefore, the average total resource consumption of the
whole network is expressed as:
\begin{equation}
	\begin{aligned}\label{EQ_Wtot}
W_{\rm tot} = \underbrace{H_w R \sum_{n=1}^N { \alpha_n^{-1}}}_{\text{multicast component}} \,+\,  \underbrace{W^{\rm UC}(u,\lambda_u) \mathcal{O}^{\rm MC}\left( \boldsymbol{p},\boldsymbol{\alpha} \right)}_{\text{unicast component}}.
\end{aligned}
\end{equation}
Accordingly, we can obtain the service outage probability, 
defined as the probability that a typical UE being served by the
hybrid scheme is in outage. It depends on the outage of the SPUC and
OMPMC components as:
$$
\mathcal{O}_{\rm tot} = \mathcal{O}^{\rm UC}(u,\gamma_{\rm th})\, \mathcal{O}^{\rm MC}(\boldsymbol{p},\boldsymbol{\alpha}).
$$

%/\/\/\/\/\/\/\/\/\/\/\/\/\/\/\/\/\/\/\/\/\/\/\/\/\/\/\/\/\/\/\/\/\/\/\/\/\/\/\/\/\/\/\/\/\/\/\/\/\/\/\/\/\/\/\/%
%  													   	 Sec_Optimizatoin										%
%/\/\/\/\/\/\/\/\/\/\/\/\/\/\/\/\/\/\/\/\/\/\/\/\/\/\/\/\/\/\/\/\/\/\/\/\/\/\/\/\/\/\/\/\/\/\/\/\/\/\/\/\/\/\/\/%
\section{Optimal Traffic Offloading for Hybrid Scheme}
\label{Sec_Optimizatoin}
Aiming to minimize the total resources consumed by the network,
the traffic offloading policy is optimized as:
\begingroup\makeatletter\def\f@size{9}\check@mathfonts
\begin{align}
	P_1:~ &\min_{\boldsymbol{p},\boldsymbol{\beta},u} ~ 
	H_wR\bsum_{n=1}^N  {\beta_n} + {W^{\rm UC}(u,\lambda_u)} \bsum_{n=1}^N f_n \mathcal{O}_n^{\rm MC}\left( p_n,\Big\{\frac{1}{\beta_l}\Big\}_l \right) \nonumber \\ \nonumber
	&~~\mbox{s.t.} 
	\begin{cases}
		\beta_n \geq 0,  &0\leq p_n\leq 1,\qquad\quad n \in S_N,\\		
		\bsum_{n=1}^N p_n \leq M, &u \in \{1,\ldots,L\},
	\end{cases}
\end{align}
\endgroup
where $\beta_n=\frac{1}{\alpha_n}$ and $S_N = \{1,\ldots,N\}$. 

Note that $P_1$ is a non-convex mixed-integer optimization problem.
The optimization parameters are the cache weights and resource
allocations of the multicast component, i.e., $\{(p_n,\beta_n)\}_1^N$,
and the multiuser beamforming parameter $u$ of the unicast component .

To optimize $P_1$, w.r.t. $u$, we use a simple line search with $u \in \{1,\ldots,L\}$.
However, to find the optimum value for $\{(p_n,\beta_n)\}_1^N$, 
we use a path-following method \cite{Vyacheslav_2014}.
%to sequentially find the optimal solution.
%We show that this approach satisfies the KKT conditions and hold the Property 3, which can guarantee a global solution
%based on Proposition \ref{Propos_GlobalSpec}.
As such, we formulate a parametric optimization problem exactly as
$P_1$ but with $\theta$-parameterized popularity $b_n(\theta)$ replacing $f_n$. 
We denote the corresponding parametric
optimization problem by $P_1(\theta)$. We parameterize
$b_n(\theta)$ such that $\lim_{\theta \rightarrow 0} b_n(\theta) =
\frac{1}{N}$ and $\lim_{\theta \rightarrow \tau} b_n(\theta) = f_n$,
for $n\in S_N$. As such, the solution of $P_1(\theta)$ can be found by
solving ODE \eqref{EQ_ODE}, written at top of next page, where $i,j
\in S_N$, $\delta_{ij}$ is the Kronecker delta function, $[a_{ij}
]_{i,j}$ constitutes a matrix whose $i$-th row and $j$-th column is
$a_{ij}$ and $[ a_{i} ]_{i} $ is a column vector with i-th component
being $a_i$.
\begin{figure*}[t]
	\begingroup\makeatletter\def\f@size{9}\check@mathfonts
	\begin{equation}\label{EQ_ODE}
		\begin{aligned}
			\begin{pmatrix}
				\dot{\boldsymbol{\beta}}(\theta) \\ \dot{\boldsymbol{p}}(\theta) \\v(\theta)
			\end{pmatrix} =-
			\begin{pmatrix}
				\left[ b_i(\theta) \dfrac{d^2 \mathcal{O}^{\rm MC}_i}{d p_i d\beta_j} \right]_{i,j} 	& 
				\left[b_i(\theta) \dfrac{d^2 \mathcal{O}^{\rm MC}_i}{d p_i^2}\delta_{ij} \right]_{i,j}  & 
				\boldsymbol{1}\vspace{5 pt}\\ 
				\left[ \mathlarger{\sum}_{n=1}^N b_n(\theta) \dfrac{d^2 \mathcal{O}^{\rm MC}_n} {d \beta_i d\beta_j}\right]_{i,j} 	& 
				\left[ b_j(\theta) \dfrac{d^2 \mathcal{O}^{\rm MC}_j}{d p_j d\beta_i} \right]_{i,j}  &  
				\boldsymbol{0}\vspace{5 pt}\\
				\boldsymbol{0}^\top & \boldsymbol{1}^\top & 0
			\end{pmatrix}^{-1}
			\begin{pmatrix}
				\left[ \dot{b}_i(\theta) \dfrac{d \mathcal{O}^{\rm MC}_i}{d p_i}\right]_{i} \vspace{5 pt}\\ 
				\left[  \mathlarger{\sum}_{n=1}^N \dot{b}_n(\theta) \dfrac{d \mathcal{O}^{\rm MC}_n}{d \beta_i} \right]_{i} \vspace{5 pt} \\ 
				0
			\end{pmatrix} 
		\end{aligned}
	\end{equation}
\endgroup
\hrule height 0.1pt depth 0pt width 7.15in \relax
\end{figure*}
We solve ODE \eqref{EQ_ODE} for $\theta \in [0,\tau]$ by the sequential method elaborated in \cite{Ours_2021}.
It gives the solution of the sough optimization problem $P_1(\tau)$
of interest, corresponding to the target popularities $\{f_n\}_1^N$.
However, an initial point, at $\theta=0$, is needed for solving this ODE.
To obtain it, we consider the optimization problem $P_1(0)$ corresponding to the popularity $b_n(0)$
and find its optimal solution.
%It can be simply obtained by an one-dimensional search considering this fact that $b_n(0)$ is independent of $n$. 
For the parameterization, we set $b_n(\theta) = {n^{-\theta}}/{\sum_{m=1}^N m^{-\theta}}$ 
for $n\in S_N$ and $\theta \in [0,\tau]$.
%Note that the solution of this approach satisfies \eqref{EQ_KKT_1}-\eqref{EQ_KKT_3}. 
%Further, based on the numerical evaluations they are decreasing w.r.t. $n$. 
%Therefore, based on Proposition \ref{Propos_GlobalSpec}, the global solutions are obtained.

%/\/\/\/\/\/\/\/\/\/\/\/\/\/\/\/\/\/\/\/\/\/\/\/\/\/\/\/\/\/\/\/\/\/\/\/\/\/\/\/\/\/\/\/\/\/\/\/\/\/\/\/\/\/\/\/%
%  													   	 Sec_Result 											%
%/\/\/\/\/\/\/\/\/\/\/\/\/\/\/\/\/\/\/\/\/\/\/\/\/\/\/\/\/\/\/\/\/\/\/\/\/\/\/\/\/\/\/\/\/\/\/\/\/\/\/\/\/\/\/\/%
\section{Simulation Results and Discussion}\label{Sec_Result}

We compare the optimal cache delivery solution of the hybrid scheme 
to conventional multi-antenna SPUC \cite{Chen2016,Wu2017,Liu2017} 
and OMPMC schemes \cite{Ours_2020} from the literature.
 
We consider the following scenario. The number of files is $N=100$,
the popularity skewness is $\theta=0.6$, the cache capacity of BSs is
$M=10$, the streaming service rate $R = 1$ Mbps,
%\in [0.1,  10]$ Mbps,
and having $L=8$ antennas at the BSs and the service
threshold $\gamma_{\rm th}=0.1$, unless it is specified. We set the
Harmonic number $H_w=6$, which approximately reduces the streaming
latency with a factor of 226~\cite{Juhn1997}. The BSs and HNs are
deployed based on two independent homogeneous PPP with intensity
$\lambda_b = 200$ and $\lambda_h\in [20,200]$, respectively, and UEs
are located according to another homogeneous PPP with intensity
$\lambda_{u} = 2\times 10^5$. We apply an Urban NLOS scenario from 3GPP
\cite{3GPP2017} with carrier frequency 2 GHz, HN transmission power 23
dBm, and path-loss exponent $e=3.76$. 
The network of BSs is
assumed interference-limited. The antenna gain at the UE and HN
are 0 dBi and 8 dBi, respectively, the noise-figure of UE is 9
dB, the noise spectrum density is -174 dBm. Note that since the
reference distance is 1 km, the UE and BS intensities are in the units
of points/km$^2$.

\begin{figure}[t]
	\begin{center}
		\includegraphics[width=85 mm]{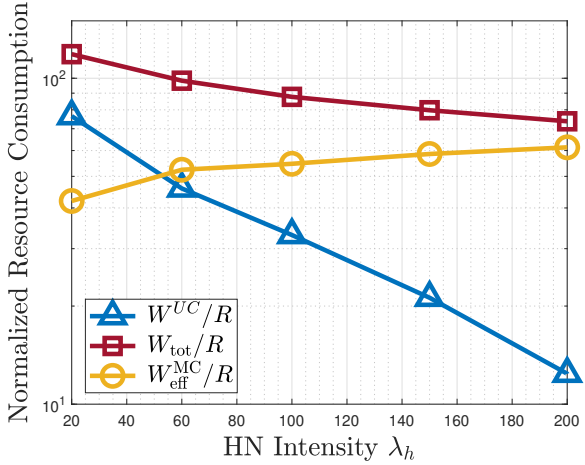}
		\caption{The normalized resource consumption as a function of HN intensity $\lambda_h$ for $\lambda_b=200$. \label{Fig_Resource_Lamp}} 
	\end{center}
\end{figure}

Figure \ref{Fig_Resource_Lamp} shows the normalized resource usage for the
hybrid scheme, and the SPUC and OMPMC components as a function of HN
intensity $\lambda_h$. % with $\lambda_u=10^5$, $\lambda_b=200$.
%Note that the resources of the SPUC component is insensitive to
%$\lambda_h$.
As $\lambda_h$ increases, the total resource consumption
decreases--the OMPMC component consumes more resources, as more files
are offloaded to be served in OMPMC. 

\begin{figure}[t]
	\begin{center}
		\includegraphics[width=85 mm]{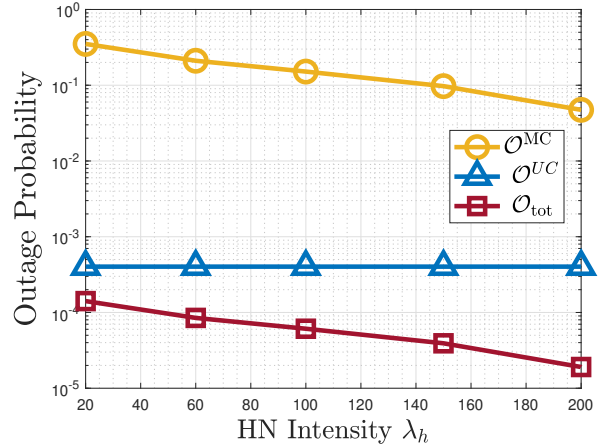}
		\caption{The outage probability as a function of HN intensity $\lambda_h$ for $\lambda_b=200$. \label{Fig_Outage_Lamp}} 
	\end{center}
\end{figure}

Figure \ref{Fig_Outage_Lamp} illustrates the outage probability as a
function of HN intensity $\lambda_h$. % for $\lambda_u=10^5$, $\lambda_b=200$.
Note that the outage probability of the SPUC
component is insensitive to $\lambda_h$. As $\lambda_h$ increases, the
total outage probability of the hybrid scheme decreases due to the
increase in the performance of the OMPMC component.

%Figure \ref{Fig_Resource_Rate} shows consumed resources 
%as a function of streaming rate $R$.
%% for $\lambda_u=10^5$, $\lambda_b= \lambda_h=200$.
%As $R$ grows, the consumed resource for the hybrid scheme, SPUC and OMPMC compo%nents, interestingly increases
%with an approximately linear scale, though \eqref{EQ_Wtot} does not
%directly indicate this fact.


\begin{figure}[t]
	\begin{center}
		\includegraphics[width=85 mm]{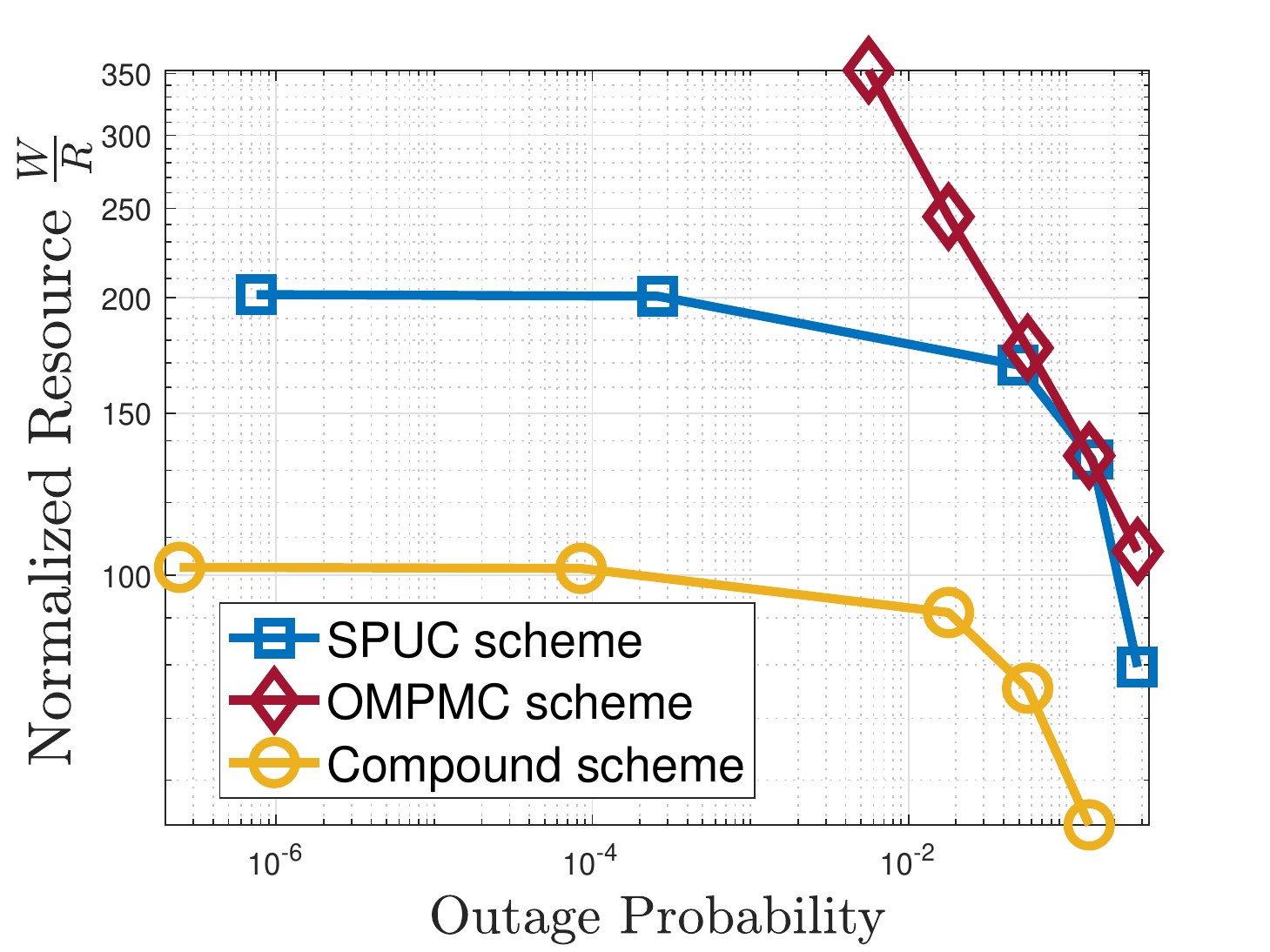}
		\caption{Normalized resource consumption versus outage probability for   $\lambda_b=200$, $\lambda_h=50$. \label{Fig_Tradeoff1}} 
	\end{center}
\end{figure}

To investigate how the proposed hybrid scheme optimizes the total
resource usage of the network, we compare it with OMPMC and SPUC
delivery schemes. Figure \ref{Fig_Tradeoff1} portrays the tradeoff
between total resource usage and outage probability of the different
schemes for $\lambda_b=200$, and a sparse MPMC component with $\lambda_h=50$. 
To generate the curves of SPUC and
hybrid schemes, we change the service threshold in the range
$\gamma_{\rm th} \in [0.03,1]$. For the OMPMC scheme, we target a
value for the outage probability and determine the corresponding
consumed resources. The OMPMC scheme performs worst as compared to the
SPUC and the hybrid schemes. The reason for this is that the considered Zipf
distribution with $\tau=0.6$ has a fat tail---there is a significant
number of request for non-cached files. Despite this, the hybrid
scheme is able to considerably outperform the SPUC scheme, offloading
the most popular files to the OMPMC component.

\begin{figure}[t!]
	\begin{center}
		\includegraphics[width=85 mm]{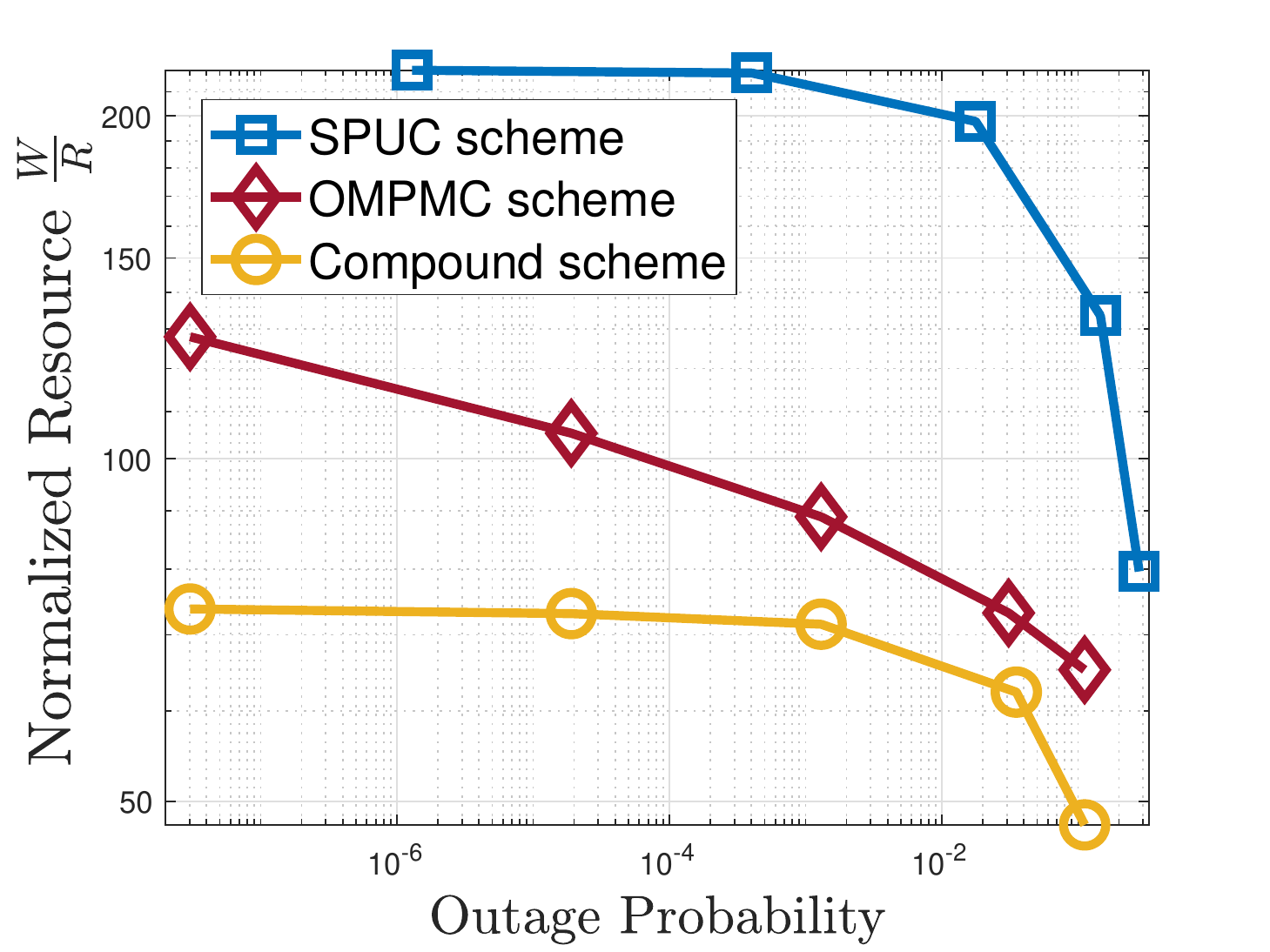}
		\caption{The  normalized resource versus outage probability for  $\lambda_b=\lambda_h=200$. \label{Fig_Tradeoff2}} 
	\end{center}
\end{figure}

The same tradeoff for a cellular network, where
$\lambda_b=\lambda_h=200$, is depicted in Figure \ref{Fig_Tradeoff2}.
In this case, the BS and caching HNs can be considered to be the same.
Now, OMPMC generically outperforms SPUC, but the hybrid scheme still
provides the best service with a wide margin. 
%
%These results indicate the merit of hybrid scheme against SPUC and
%OMPMC ones and show that employing the proposed scheme has positive
%effects on the system performance in terms of network-wide consumed
%resource and outage probability.
%\begin{figure}[t]
%	\begin{center}
%		\includegraphics[width=90 mm]{Resource_Rate.eps}
%		\caption{Consumed resources as a function of rate $R$ for $\lambda_u=10^5$, $\lambda_b= \lambda_h=200$. \label{Fig_Resource_Rate}} 
%	\end{center}
%\end{figure}
%\begin{figure}[t!]
%	\begin{center}
%		\includegraphics[width=95 mm]{Outage_Rate.eps}
%		\caption{The outage probability as a function of rate $R$ for $\lambda_u=10^5$, $\lambda_b= \lambda_h=200$. %$\lambda_b=\lambda_b=200$. \label{Fig_Outage_Rate}} 
%	\end{center}
%\end{figure}

%
%Figure \ref{Fig_alpha} shows the spectral efficiency of traffic-offloading policy 
%as a function of the file index for UE intensity $\lambda_u=10^5$, 
%BS cache intensity $\lambda_b=200$ and service threshold $\gamma_{\rm th}=0.03$.
%According to Property 4, the values not being sketched in this figure have infinite spectral %efficiency values
%(or correspondingly zero values for their $\beta$).
%The solutions are increasing w.r.t. $n$ for all evaluated intensity $\lambda_h$, as declared in %Property 3.
%According to Proposition \ref{Propos_GlobalSpec}, 
%it shows that the used solution approach has been able to find the global optimum.
%As HN intensity increases, the classifier index $K$ and spectral efficiency $\alpha_n$ grow,
%which shows a trade-off between $K$ and OMPMC file-specific resource $w_n^{\rm MC}$.
%This shows that as HN intensity increases, the OMPMC component needs to stream more files.
%\begin{figure}[t!]
%	\begin{center}
%		\includegraphics[width=90 mm]{alpha.eps}
%	\caption{The bandwidth allocation as a function of file index for $\lambda_u=10^5$, $\lambda_b=200$ 
%		and $\gamma_{\rm th}=0.03$. \label{Fig_alpha}} 
%	\end{center}
%\end{figure}

%/\/\/\/\/\/\/\/\/\/\/\/\/\/\/\/\/\/\/\/\/\/\/\/\/\/\/\/\/\/\/\/\/\/\/\/\/\/\/\/\/\/\/\/\/\/\/\/\/\/\/\/\/\/\/\/%
%  													   	 Conclusoin 											%
%/\/\/\/\/\/\/\/\/\/\/\/\/\/\/\/\/\/\/\/\/\/\/\/\/\/\/\/\/\/\/\/\/\/\/\/\/\/\/\/\/\/\/\/\/\/\/\/\/\/\/\/\/\/\/\/%
\section{Conclusion}\label{Sec_Conclu}

In this paper, we considered hybrid content delivery combining
orthogonal multipoint multicast (OMPMC) and single-point unicast
(SPUC) transmission schemes. A traffic offloading policy jointly
optimizing cache placement and radio resource allocation of OMPMC and
multiuser beamforming of SPUC was formulated, based on the derived
expression of network-wide consumed resource. The amount of consumed
resources depends on the ratio of user intensity to SPUC BS intensity,
and not on both separately. Further, a functional relationship between
the total resource consumption and the outage probability of unicast
service was derived. To find the optimal policy, a parametric
optimization problem was formulated and solved using a path-following
method.
%The solution characteristic showed that the files are classified into two sets, popular and less popular ones,
%where the less popular set is not served by OMPMC,
%while popular set is served by both OMPMC and SPUC components.
We compared the performance of the hybrid scheme with using only SPUC
or OMPMC. Simulation results clearly portray the tradeoff between total
resource consumption and service outage probability, and showed that
despite a fat tail of popularity distribution, which renders OMPMC
ineffective as compared to SPUC, the hybrid scheme outperforms SPUC
with  a wide margin, in the simulated scenarios up to  50\% of the
resources can be saved by using the hybrid scheme instead of SPUC. 
%delivery strategies from the total resource of network and service
%outage probability perspectives and for different values of cache
%intensity, reliable rate and service threshold.
The proposed hybrid delivery scheme is a promising candidate for
optimizing the spectral efficiency and total resource consumption in
heterogeneous and cellular networks.

\section*{Acknowledgment}
This work was funded in part by the Academy of Finland (grant 319058).
The work of  G. Caire was partially funded by the European Research Council under the ERC Advanced Grant N. 789190, CARENET.

%\small{
\bibliographystyle{IEEEtran}
\bibliography{IEEEabrv,IEEE}

%/\/\/\/\/\/\/\/\/\/\/\/\/\/\/\/\/\/\/\/\/\/\/\/\/\/\/\/\/\/\/\/\/\/\/\/\/\/\/\/\/\/\/\/\/\/\/\/\/\/\/\/\/\/\/\/%
%  													   	 Appendix   											%
%/\/\/\/\/\/\/\/\/\/\/\/\/\/\/\/\/\/\/\/\/\/\/\/\/\/\/\/\/\/\/\/\/\/\/\/\/\/\/\/\/\/\/\/\/\/\/\/\/\/\/\/\/\/\/\/%

\appendix
\subsection{Proof of Theorem \ref{Theorem_Wt}}\label{App1}
Before going through proof of Theorem \ref{Theorem_Wt}, we need the following Lemma.
\begin{lem}\label{Lemma1}
	Let $\{\boldsymbol{x}_i\}_i$ be the points of a homogeneous PPP $\Phi$ with intensity $\lambda$,
	and $S(\cdot)$ and $P(\cdot)$ be two real-valued functions on the state space of $\Phi$, 
	then we have:
	\begin{align*}
		\mathbb{E}\left\{ \prod_{k\in \Phi} P(\boldsymbol{x}_k) \sum_{k\in \Phi} S(\boldsymbol{x}_k) \right \} 
		&= \lambda \iint_{\mathbb{R}^2} S(\boldsymbol{s})P(\boldsymbol{s}) d\boldsymbol{s} \times  \\
		&\exp\left( \lambda \iint_{\mathbb{R}^2} \big( P(\boldsymbol{s}) - 1\big)d\boldsymbol{s} \right),
	\end{align*}
	where $\boldsymbol{s} = (x,y)$	and $d\boldsymbol{s}=dxdy$.
	\begin{proof}
		Let $N$ denotes the number of points of $\Phi$, which follows a Poisson distribution.
		Suppose that the points are placed in the region $A$ as a subspace of Cartesian space.
		Note that for a homogeneous PPP, the points can be considered to be independently and uniformly distributed 
		over $A$ \cite{Chiu2013}. 
		We then have:
		\begingroup\makeatletter\def\f@size{8.5}\check@mathfonts
		\begin{align*}
			&\mathbb{E}\Big\{ \prod_{k\in \Phi} P(\boldsymbol{x}_k) \sum_{k\in \Phi} S(\boldsymbol{x}_k) \Big \} =
			\mathbb{E}_N \bigg\{ \mathbb{E} \Big\{\sum_{k=1}^N S(\boldsymbol{x}_k)P(\boldsymbol{x}_k)\prod_{i\ne k}^N P(\boldsymbol{x}_i) \Big| N \Big\} \bigg\}\\
			&\quad\overset{(a)}= \mathbb{E}_N\left\{ \sum_{n=1}^N \iint_A \frac{S(\boldsymbol{s})P(\boldsymbol{s})}{|A|}d\boldsymbol{s} \: \Big(\iint_A \frac{P(\boldsymbol{s})}{|A|} d\boldsymbol{s}\Big)^{N-1}  \right\} \\
			&\quad= \iint_A \frac{S(\boldsymbol{s})P(\boldsymbol{s})}{|A|}d\boldsymbol{s} ~\mathbb{E}_N\left\{ N  \: \Big(\iint_A \frac{P(\boldsymbol{s})}{|A|} d\boldsymbol{s}\Big)^{N-1} \right\} \\
			&\quad\overset{(b)}=  \iint_A \frac{S(\boldsymbol{s})P(\boldsymbol{s})}{|A|}d\boldsymbol{s} \: \lambda |A| \exp\big(-\lambda|A| \big) \exp\left(\lambda |A| \iint_A \frac{P(\boldsymbol{s})}{|A|}d\boldsymbol{s}\right)
		\end{align*}
		\endgroup
		where $|A|$ is the area of $A$, for $(a)$ we consider that points are distributed over $A$ 
		with $dS=dxdy$,
		and for $(b)$ we consider this fact that $N\sim \mbox{Pois}(\lambda |A|)$.
		By rearranging the recent equation and letting $A=\mathbb{R}^2$, statement follows.
	\end{proof}
\end{lem}
We also need to compute the expectation of resource needed for a typical UE requesting 
from the SPUC component. 
Based on \eqref{EQ_SINR} and \eqref{EQ_serviceTh} and by defining $w_i = \dfrac{R}{\log(1+\gamma_i)}$, we have:
\begingroup\makeatletter\def\f@size{9}\check@mathfonts
\begin{align*}
	\mathbb{E}_{ {\gamma}_i }\left\{  \frac{R}{\log(1+\gamma_i)} \mathbbm{1}(\gamma_i \geq \gamma_{\rm th})\right\} 
	&= \mathbb{E}_{ w_i }\bigg\{  \frac{R}{\log(1+\gamma_i)}  \mathbbm{1}(w_i \leq w_{\rm th})\bigg\} \\
	&= \int_0^{w_{\rm th}} w \frac{d}{dw} \mbox{F}_{w_i}(w) dw,
\end{align*}
\endgroup
where $\mbox{F}_{w_i}(w)$ is the CDF of r.v. $w_i$ evaluated at $w$.
Based on \eqref{EQ_SINR}, we obtain:
\begin{align*}
	\mbox{F}_{w_i}(w) &= 1 - \mathbb{P}\left\{ g_i\|\boldsymbol{x}_i\|^{-e}\leq I_i \eta(w) \right\} \\
	&\overset{(a)}= \sum_{l=1}^{L-u+1}f_{l,u} \mathbb{E}_{I_i}\left\{ \exp(-l\xi \eta(w) I_i \|\boldsymbol{x}_i \|^{e}) \right\},
\end{align*}
where $(a)$ obtained based on the gamma distribution of $g_i \sim \Gamma(L-u+1,1)$ 
and promising approximation used in \cite{Xu2019}, though for $L=1$ it is exact.
We also have $f_{l,u} =  (-1)^{l+1}{L-u+1 \choose l}$ and $\eta(w) = 2^{\frac{R}{w}-1}$.
Considering that $I_i = \sum_{k \in \Phi_m} g_k^i \|\boldsymbol{x}_i-\boldsymbol{r}_k \|^{-e}$ 
and $g_k^i\sim\Gamma(u,1)$ we can get:
\begingroup\makeatletter\def\f@size{9}\check@mathfonts
\begin{align*}
	\mathbb{E}_{I_i}\left\{ \exp(-l\xi \eta(w) I_i \|\boldsymbol{x}_i \|^{e}) \right\} =& \prod_{k\in \Phi_m} \left( 1+l\xi\eta(w) \frac{\|\boldsymbol{x}_i-\boldsymbol{r}_k\|^{-e}}{\|\boldsymbol{x}_i\|^{-e}} \right)^{-u} \\
	=& \prod_{k\in \Phi_m} P_{w,l}\left(\frac{\|\boldsymbol{x}_i-\boldsymbol{r}_k \|^{-e}}{\|\boldsymbol{x}_i\|^{-e} }\right),
\end{align*}
\endgroup
where $~\: P_{w,l}(\zeta ) := \big( 1+l\xi\eta(w)\: \zeta  \big)^{-u}$.
By using $\dfrac{d}{dw}\mbox{F}_{w_i}(w) = \mbox{F}_{w_i}(w) \dfrac{d}{dw}\:\log\left( \mbox{F}_{w_i}(w)\right)$, 
and defining 
$$
S_{w,l}(\zeta) := -u \frac{d\eta(w)}{dw} \dfrac{l\xi\:\zeta}{1+l\xi\eta(w)\:\zeta},
$$ we can obtain:
\begin{align}\label{EQ_Exp_w}
	&\mathbb{E}_{ {\gamma}_i  }\left\{  \frac{R}{\log(1+\gamma_i)} \mathbbm{1}(\gamma_i \geq \gamma_{\rm th})\right\} 
	= \sum_{l=1}^{L-u+1}f_{l,u} \int_0^{w_{\rm th}} w \:\times \notag\\
	&\quad\sum_{k\in\Phi_m} S_{w,l} \left(\frac{\|\boldsymbol{x}_i-\boldsymbol{r}_k \|^{-e}}{\|\boldsymbol{x}_i\|^{-e} }\right) 
	\prod_{k\in \Phi_m} P_{w,l}\left(\frac{\|\boldsymbol{x}_i-\boldsymbol{r}_k \|^{-e}}{\|\boldsymbol{x}_i\|^{-e} }\right) dw.
\end{align}

Now, we can compute $W' = \mathbb{E}\left\{ \sum_{i \in \mathcal{V}_0} w_i \right\}$ from \eqref{EQ_WUC0}.
Without loss of generality, we assume that BS responsible in the cell $\mathcal{V}_0$ is located at origin, 
i.e., $\boldsymbol{r}_0=\boldsymbol{0}$. 
Based on \eqref{EQ_SINR}, we have:
\begingroup\makeatletter\def\f@size{9}\check@mathfonts
\begin{equation}\label{EQ_wt1}
	\begin{aligned}
		& W' = \mathbb{E}_{\Phi_u,\Phi_m} \mathbb{E}_{ \boldsymbol{\gamma} } \left\{  \sum_{i \in \mathcal{V}_0} \frac{R}{\log(1+\gamma_i)} \mathbbm{1}(\gamma_i \geq \gamma_{\rm th}) \right\} \\
		&= \mathbb{E}_{\Phi_u,\Phi_m} \left\{  \bsum_{i \in \Phi_u}  \mathbb{E}_{ \gamma_i }\left\{  \frac{R}{\log(1+\gamma_i)} \mathbbm{1}(\gamma_i \geq \gamma_{\rm th})\right\} \mathbbm{1}(\boldsymbol{x}_i \in \mathcal{V}_0)   \right\} \\
		&\overset{(a)}=  \mathbb{E}_{\Phi_u}\bigg\{ \bsum_{i \in \Phi_u}  \mathbb{E}_{\Phi_m}\Big\{ \bsum_l f_{l,u} \int_0^{w_{\rm th}} w\bsum_{k\in \Phi_m \backslash \{0\}} S_{w,l}\left(\frac{\|\boldsymbol{x}_i-\boldsymbol{r}_k \|^{-e}}{\|\boldsymbol{x}_i\|^{-e} }\right)  \\
		&\qquad \prod_{k\in \Phi_m \backslash \{0\}} P_{w,l}\left(\frac{\|\boldsymbol{x}_i-\boldsymbol{r}_k \|^{-e}}{\|\boldsymbol{x}_i\|^{-e} }\right) \mathbbm{1}\big(\boldsymbol{x}_i \in M_0(\boldsymbol{r}_k)\big) dw \Big\} \bigg\}\\
	\end{aligned}
\end{equation}
\endgroup
where for $(a)$ we use \eqref{EQ_Exp_w} 
and the property from the Voronoi tessellation that $\mathbbm{1}(\boldsymbol{x}_i \in \mathcal{V}_0)$ can be expressed 
based on a product of Poisson points of $\Phi_m$:
$$
\mathbbm{1}(\boldsymbol{x}_i \in \mathcal{V}_0) = \prod_{k\in \Phi_m \backslash \{0\} } \mathbbm{1}\big( \boldsymbol{x}_i \in M_0\left(\boldsymbol{r}_k\right) \big),
$$
where $M_0(\boldsymbol{r}_k) = \big\{\boldsymbol{r}'\in \mathbb{R}^2 ~\big|~ \|\boldsymbol{r}'-\boldsymbol{r}_0\| \leq \|\boldsymbol{r}'-\boldsymbol{r}_k\| \big\}$ for $k \in \Phi_m \backslash \{0\}$.
%as plotted in Figure \ref{?}.
\begin{figure*}[t]
	\begin{equation}\label{EQ_app1}
		\begin{aligned}
			W' =&  \bsum_{l=1}^{L-u+1} f_{l,u} \int_0^{w_{\rm th}} w \lambda_b \lambda_u \iint_D \Bigg\{ \iint_{B_c( \boldsymbol{x},\|\boldsymbol{x}\| )} Q_{w,l}\left(\frac{\|\boldsymbol{x}-\boldsymbol{r} \|^{-e}}{\|\boldsymbol{x}\|^{-e} }\right) rdrd\theta \\ 
			&\exp\Big(\lambda_b \iint_{B_c( \boldsymbol{x},\|\boldsymbol{x}\| )} P_{w,l}\left(\frac{\|\boldsymbol{x}-\boldsymbol{r} \|^{-e}}{\|\boldsymbol{x}\|^{-e} }\right) rdrd\theta - \lambda_b\iint_D rdrd\theta  \Big) \Bigg\}xdxd\theta'\:dw
		\end{aligned}
	\end{equation}
	\hrule height 0.1pt depth 0pt width 7.15in \relax
\end{figure*}
Note that in \eqref{EQ_wt1} we are dealing with a reduced-palm process as $k\in\Phi_m \backslash \{0\}$.
Taking into account that the distribution of reduced-palm process is equal 
to the distribution of original process \cite{BK1}, 
and based on Lemma \ref{Lemma1}, we have:
\begingroup\makeatletter\def\f@size{9}\check@mathfonts
\begin{align*}
	W' &=  \bsum_{l=1}^{L-u+1} f_{l,u} \int_0^{w_{\rm th}} w \mathbb{E}_{\Phi_u}\bigg\{ \sum_{i\in\Phi_u} \lambda_b\iint_D Q_{w,l}\left(\frac{\|\boldsymbol{x}_i-\boldsymbol{r} \|^{-e}}{\|\boldsymbol{x}_i\|^{-e} }\right) \\
	&\quad\mathbbm{1}\big(\boldsymbol{x}_i\in M_0(\boldsymbol{r})\big)rdrd\theta \: \exp\bigg( \lambda_b \iint_D \Big(P_{w,l}\left(\frac{\|\boldsymbol{x}_i-\boldsymbol{r} \|^{-e}}{\|\boldsymbol{x}_i\|^{-e} }\right) \\
	&\quad \mathbbm{1}\big(\boldsymbol{x}_i\in M_0(\boldsymbol{r})\big)-1\Big) rdrd\theta \bigg) \bigg\}dw,
\end{align*}
\endgroup
where $D=\{(r,\theta)|\:0\leq r \leq \infty, 0 \leq \theta \leq 2\pi\}$ and $Q_{w,l}(\cdot):=S_{w,l}(\cdot)P_{w,l}(\cdot)$.
By considering:
$$
\mathbbm{1}\big(\boldsymbol{x}_i\in M_0(\boldsymbol{r})\big) = B_c\big( \boldsymbol{x}_i,\|\boldsymbol{x}_i\| \big),
$$
where $B_c(\boldsymbol{x},\|\boldsymbol{x}\|)$ denotes the region outside 
the circle located at $\boldsymbol{x}$ with radius $\|\boldsymbol{x}\|$, we get \eqref{EQ_app1},
where $x=\| \boldsymbol{x}\|$.
Note that \eqref{EQ_app1} obtained by considering the Probability-Generation-Functional (PFGL) property of PPP $\Phi_u$ \cite{BK1}.
Now, we consider change of variables $\boldsymbol{\rho} = \boldsymbol{x}-\boldsymbol{r}$, 
and then $\boldsymbol{q} = z\boldsymbol{\rho}$ to obtain:
\begingroup\makeatletter\def\f@size{9}\check@mathfonts
\begin{align*}
	W' =&  \bsum_{l=1}^{L-u+1} f_{l,u} \int_0^{w_{\rm th}} w \lambda_b \lambda_u \iint_D \Big\{ 2\pi \int_1^\infty Q_{w,l}\left( z^{-e}  \right) q^2 zdz\\ 
	&\exp\Big( 2\pi\lambda_b q^2 \big(\int_1^\infty P_{w,l}\left( z^{-e} \right)zdz - \int_0^\infty zdz\big)  \Big) \Big\}qdqdw \\[4 pt]
	\overset{(a)}=& \bsum_{l=1}^{L-u+1} f_{l,u} \int_0^{w_{\rm th}}  \Big( \frac{w\lambda_u}{2\lambda_b}\int_1^\infty Q_{w,l}(z^{-e})zdz \:\times \\
	&\qquad\qquad \qquad \qquad \quad \dfrac{1}{\left(\frac12-\int_1^\infty \left(P_{w,l}(z^{-e})-1\right)zdz \right)^2} \Big)dw \\[5 pt]
	=& \frac{\lambda_u}{2\lambda_b} \bsum_{l=1}^{L-u+1} f_{l,u} \int_0^{w_{\rm th}} w \frac{d}{dw}\Big(\frac{1}{\frac12-\int_1^\infty \left(P_{w,l}(z^{-e})-1\right)zdz }\Big)dw		
\end{align*}
\endgroup
where $\rho=\| \boldsymbol{\rho}\|$, $q=\| \boldsymbol{q}\|$, $\dfrac{d}{dw}P_{w,l}(\cdot)=Q_{w,l}(\cdot)$ 
and for $(a)$ we use $\int_0^\infty q^3 \exp(\phi_0 q^2) dq = \dfrac{1}{2\phi_0^2}$ for $\phi_0<0$.
By computing the inner integral, the statement follows.

To compute $\mathcal{O}^{\rm UC}(u,\gamma_{\rm th})$ as the outage probability of a typical UE served by the SPUC component,
we notice that it depends on $\mathbb{P}(\gamma_{k}^{\rm UC} \leq \gamma_{\rm th})$ and can be computed based on \eqref{EQ_serviceTh}.
As such, the same methodology as applied for $W'$ can be used.
\end{document}